\RequirePackage[2020-02-02]{latexrelease}
\documentclass[twocolumn,english,aps,prl,10pt,superscriptaddress]{revtex4}
\usepackage[T1]{fontenc}
\usepackage{amsmath,graphicx,amssymb,epsfig,dsfont,mathrsfs,color,amsbsy}
\usepackage{float}
\begin{document}

\title{Interference-Controlled Radiative Heat Transport in Time-Modulated Networks}

\author{P. Ben-Abdallah}
\email{pba@institutoptique.fr}
\affiliation{Laboratoire Charles Fabry, UMR 8501, Institut d'Optique, CNRS, Universit\'e Paris-Saclay, 2 Avenue Augustin Fresnel, 91127 Palaiseau Cedex, France}

\date{\today}


\begin{abstract}
We demonstrate photonic control of radiative heat transport in nanoscale networks through phase-controlled interference between elastic and inelastic Floquet scattering channels induced by temporal permittivity modulation. Relative modulation phases select constructive or destructive interference, enabling directional thermal-photon currents and heat splitting even at thermal equilibrium. Modulation amplitude and frequency further tune the enhancement, suppression and redistribution of energy flow. This interference-based mechanism enables thermal routing and logic operations and provides a general platform for reconfigurable photonic heat management at the nanoscale.
\end{abstract}

\maketitle

Near-field radiative heat transfer has been a central topic in thermal nanophotonics for the past two decades~\cite{Volokitin,Joulain,Cuevas,biehs1}. In this regime, energy exchange is often dominated by narrowband surface modes, such as surface phonon polaritons, which require strong spectral overlap for efficient coupling. Consequently, even modest detuning between resonances leads to a sharp suppression of heat transfer~\cite{Pendry,pba_prb_2010,Miller_prl_2015}, severely limiting the adaptability of nanoscale thermal networks. Existing approaches to overcome this limitation rely on structural redesign~\cite{Rodriguez_prl2011,Cuevas_prl_2017,Messina_prb_2017,Fan_prl_2018}, amplification mechanisms~\cite{pba_prl_2014}, or magnetic-field-induced nonreciprocity~\cite{Tang,Song,Lingling,Cuevas2,pba2016}, but remain inherently static or require large external fields.
Temporal modulation of optical properties provides a dynamic alternative, enabling control of thermal emission~\cite{Liberal,Fan1,Wang1,Wang2} and radiative heat exchange~\cite{Wang3,Wang4}. Recent experiments have demonstrated modulation at gigahertz frequencies~\cite{Hui,Chegel} and through ultrafast phononic excitation~\cite{Khalsa,Kusaba}, allowing coherent changes of infrared optical properties on sub-picosecond time scales. Using Floquet theory, directional heat flow between isothermal reservoirs has been predicted~\cite{Alcazar}, while time-dependent Green’s-function approaches have revealed enhancement, suppression and frequency conversion of radiative heat transfer~\cite{Fan2,Fan3}. A general many-body Floquet framework has recently been developped, providing a broader theoretical foundation to investigate complex driven systems~\cite{arXiv:2510.19378}. Interestingly, thermophotonic transport can be affected by intrinsic electronic orders, such as charge-density-wave transitions, which alter the band structure and optical response~\cite{Zhao}. Coupling these effects with temporal modulation could enable additional control over radiative energy transfer.

In this Letter, we show that temporal modulation of material permittivities enables tunable radiative heat transport through phase-controlled interference between elastic and inelastic Floquet scattering channels. Modulation generates $\omega_\mathrm{c}$s at $\omega\pm n\Omega$, restoring coupling between spectrally mismatched resonances, while relative modulation phases break time-reversal symmetry and produce directional heat currents even at thermal equilibrium. Interference between scattering pathways further enables dynamic redistribution of energy, allowing continuously tunable heat-flux splitting in many-body networks. These effects establish a general platform for reconfigurable photonic heat routing at the nanoscale.

We consider a network of $N$ spherical nanoparticles of radii $R_i$, permittivities $\varepsilon_i$, and temperatures $T_i$, separated by distances $r_{ij}$. In the dipolar regime $R_i\ll r_{ij}$ and $R_i\ll\lambda_i$, each particle is modeled as a point electric dipole
\begin{equation}
\mathbf{p}_i(\omega)=\varepsilon_0\,\alpha_i(\omega)\,\mathbf{E}_i^{\rm loc}(\omega),
\end{equation}
where $\alpha_i(\omega)$ is the Clausius--Mossotti polarizability. The local field reads
\begin{equation}
\mathbf{E}_i^{\rm loc}(\omega)=\mathbf{E}_i^{\rm fl}(\omega)
+\mu_0\omega^2\sum_{j\neq i}\mathbf{G}_{0,ij}(\omega)\mathbf{p}_j(\omega),
\end{equation}
with $\mathbf{G}_{0,ij}$ the vacuum dyadic Green tensor. Solving the coupled-dipole equations yields
\begin{equation}
\begin{split}
\mathbf{p}_i(\omega)
&=\varepsilon_0\alpha_i(\omega)\mathbf{E}_i^{\rm fl}(\omega)\\
&\quad+\varepsilon_0\mu_0\omega^2
\sum_{j\neq i}\alpha_i(\omega)\mathbf{G}_{0,ij}(\omega)\alpha_j(\omega)
\mathbf{E}_j^{\rm fl}(\omega)
+\cdots ,
\end{split}
\end{equation}

\textit{Matching detuned nodes.—}
In the stationary regime, the Landauer transmission coefficient between two dipoles embedded in a network reads~\cite{pba2011,messina2013}
\begin{equation}
\mathcal{T}_{ij}^{(0)}(\omega)=
4\frac{\omega^4}{c^4}
\Im[\alpha_i(\omega)]\Im[\alpha_j(\omega)]
\mathrm{Tr}\!\left[
\mathbf{G}_{ij}(\omega)
\mathbf{G}^{\dagger}_{ij}(\omega)
\right],
\end{equation}
where the full Green tensor $\mathbf{G}_{ij}$ satisfies the Dyson equation
\begin{equation}
\mathbf{G}_{ij}(\omega)=
\mathbf{G}_{0,ij}(\omega)+
\mathbf{G}_{0,ij}(\omega)\alpha_i(\omega)\mathbf{G}_{ij}(\omega).
\end{equation}
\begin{figure} [H]
	\centering
	\includegraphics[height=0.28\textwidth,angle=0]{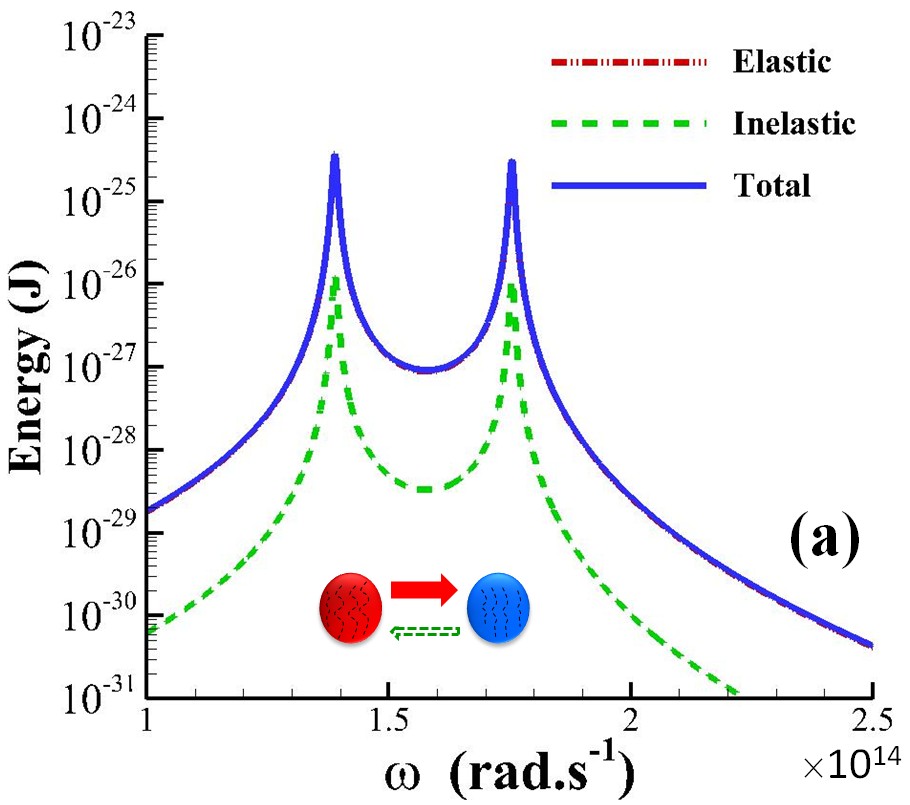}
           \includegraphics[height=0.28\textwidth,angle=0]{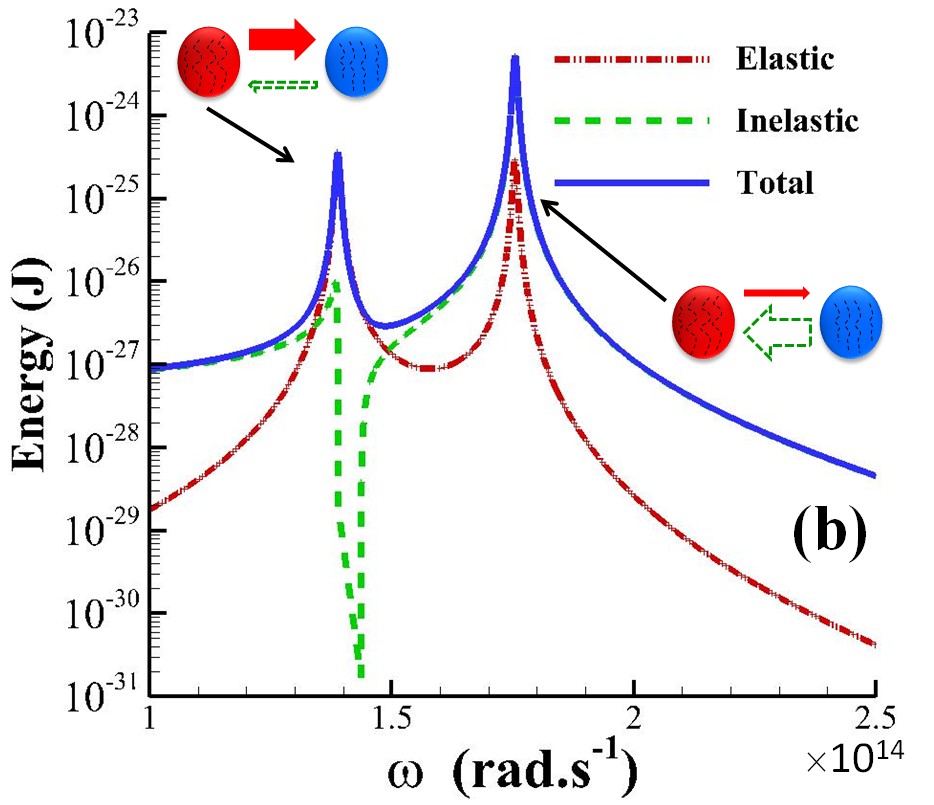}
           \includegraphics[height=0.28\textwidth,angle=0]{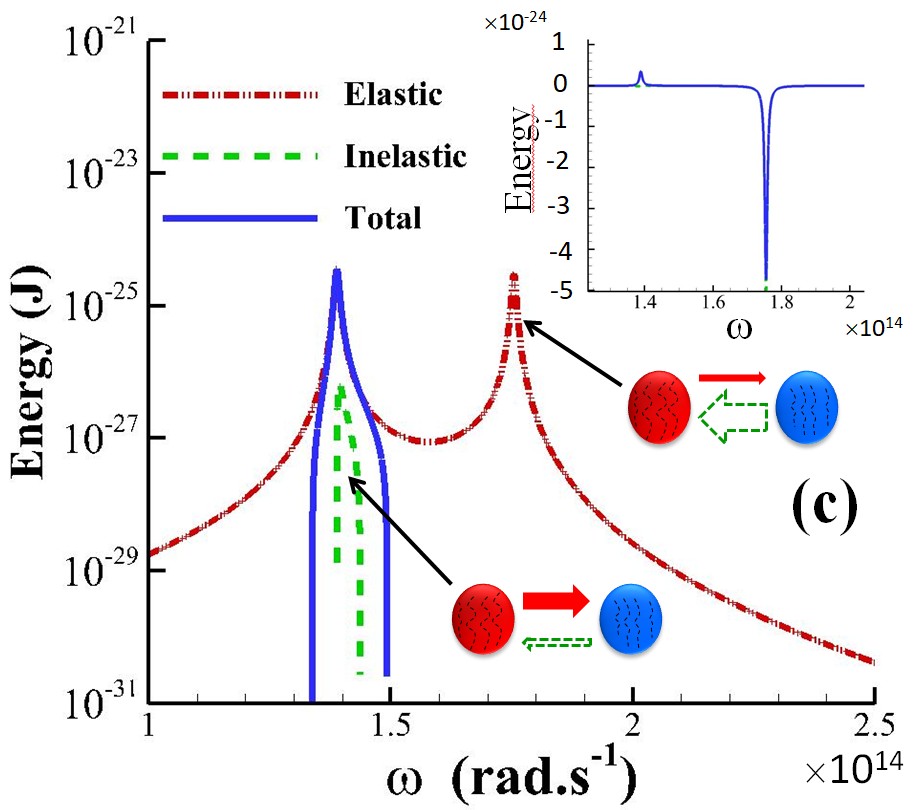}
	\caption{Out of equilibrium energy exchange between a SiC and a GaN particle ($R=50\: nm$ radius)  separated by a distance $d=3R$ when $T_{SiC}=400\,$K and $T_{GaN}=300\,$K and the particle polarizabilities undergoe a harmonic modulation  with a dephasing $\Delta \phi=\{0,\pi/2,-\pi/2\}$ in Figs (a), (b) and (c), respectively. The SiC and GaN permittivities are modeled by a Lorentz oscillator~\cite{Palik},
$\varepsilon(\omega) = \varepsilon_\infty (\omega_{\rm LO}^2 - \omega^2 - i\gamma \omega)/(\omega_{\rm TO}^2 - \omega^2 - i\gamma \omega)$,
with parameters $\varepsilon_\infty = 6.7$, $\omega_{\rm LO}= 1.825\times10^{14}$\,rad/s, $\omega_{\rm TO} = 1.494\times10^{14}$\,rad/s, $\gamma = 8.966\times10^{11}$\,rad/s for SiC, and 
$\varepsilon_\infty = 5.35$, $\omega_{\rm LO}= 1.415\times10^{14}$\,rad/s, $\omega_{\rm TO} = 1.315\times10^{14}$\,rad/s, $\gamma = 1.0\times10^{12}$\,rad/s for GaN. The modulation frequency $\Omega=\Delta\omega_{\rm LO}$. The difference in arrow thickness between the particles indicates the relative importance of elastic (red) and inelastic (green) transport processes. Inset: total and inelastic energy exchanged in linear scale.} \label{Fig1:flux matching}
\end{figure} 
If the polarizability is periodically modulated at frequency $\Omega$, it admits a Floquet expansion
\begin{equation}
\alpha_i(t)=\sum_{n=-\infty}^{\infty}\hat{\alpha}_{i,n}(\omega)e^{in\Omega t}.
\end{equation}
For harmonic modulation,
\begin{equation}
\alpha_i(t)=\alpha_{i,0}+\delta\alpha_i\cos(\Omega t+\phi_i),
\qquad |\delta\alpha_i|\ll|\alpha_{i,0}|,
\end{equation}
the only nonzero coefficients are
\begin{equation}
\hat{\alpha}_{i,0}=\alpha_{i,0},\qquad
\hat{\alpha}_{i,\pm1}=\frac{\delta\alpha_i}{2}e^{\pm i\phi_i}.
\end{equation}
In the present framework, radiative energy exchange between two nanoparticles can occur through different Floquet channels generated by the temporal modulation. The elastic channel corresponds to $n=0$, where thermal photons are exchanged without frequency conversion and therefore preserve their original frequency $\omega$. By contrast, inelastic transport arises from the sideband channels $n\neq0$, where photons exchange energy with the modulation field through the absorption or emission of quanta $\hbar\Omega$, leading to frequency-shifted exchanges at $\omega\pm n\Omega$.
To derive the corresponding transmission coefficients, we assume a weak harmonic modulation of the polarizability, $|\delta\alpha_i|\ll|\alpha_{i,0}|$, which allows the Floquet expansion to be truncated to the first-order sidebands $n=\pm1$. In this regime, the Dyson equation for the Green tensor in the Floquet basis couples only neighboring sidebands, while higher-order processes involving multiple modulation quanta can be neglected. Consequently, to leading order in the modulation amplitude, energy exchange is dominated by single-quantum absorption or emission processes.

Introducing the Landauer-like sideband transmission coefficient~\cite{arXiv:2510.19378}
\begin{equation}
\begin{split}
\mathcal{T}^{(n)}_{ij}(\omega) 
&= 4 \frac{(\omega+n\Omega)^4}{c^4} \,
\Im[\hat{\alpha}_{i,n}(\omega+n\Omega)] \\ 
&\quad \times \Im[\hat{\alpha}_{j,n}(\omega+n\Omega)] \\ 
&\quad \times \mathrm{Tr} \Big[ \mathbf{G}_{ij}(\omega+n\Omega) 
\mathbf{G}_{ij}^\dagger(\omega+n\Omega) \Big].
\end{split}
\label{Trans_inelastic}
\end{equation}

the total power exchanged from node $i$ to $j$ reads
\begin{equation}
P_{i \to j} = \sum_{n=-1}^{1} 
\int_0^\infty \frac{d\omega}{2\pi} \, 
\hbar (\omega + n\Omega) 
\Big[ n_i(\omega) - n_j(\omega + n\Omega) \Big] 
\mathcal{T}_{ij}^{(n)}(\omega).
\end{equation}
In Figs.~\ref{Fig1:flux matching} we show that the inelastic contribution to the radiative heat exchange between two detuned polar nanoparticles strongly depends on both the
relative modulation phase $\Delta\phi$ and the modulation frequency $\Omega$.
Since the first-order Floquet sidebands of each particle is proportional to
$\hat{\alpha}_{\pm1}\propto e^{\pm i\phi}$,their imaginary parts, which govern
dissipation and thus heat transfer, scale as $\Im[\hat{\alpha}_{\pm1}]\propto \sin\phi$.
A relative phase of $\Delta\phi=\pi/2$ (Fig.\ref{Fig1:flux matching}-b) thus maximizes the dissipative overlap between
the sidebands of the two particles, leading to a maximal energy flow which can be much larger than the flux in static regime. In Fig.\ref{Fig1:flux matching}-c we see also that a phase of $\Delta\phi=-\pi/2$ allows to pump heat by reversing the direction of heat flux from the cold to the hot particle. 
In Fig.~\ref{Fig1:flux matching}(a), the elastic contribution dominates by one to two orders of magnitude, so the total and elastic curves nearly overlap on the logarithmic scale. In Fig.~\ref{Fig1:flux matching}(c), the exchange is negative in some spectral regions (see linear-scale inset), so the log is shown only where it is positive, around the GaN $\omega_{\mathrm{LO}}$ resonance where elastic effects dominate. Near the SiC longitudinal optical phonon frequency, inelastic channels dominate, with the total exchange exceeding the elastic contribution by more than an order of magnitude.
Furthermore, the comparison of Figs.\ref{Fig1:flux matching} and 2 shows that the inelastic transfer is enhanced when the modulation frequency
$\Omega$ matches the difference between the longitudinal optical (LO) phonon
frequencies of the two particles, $\Omega= \omega_{\rm LO}^{(SiC)}-\omega_{\rm LO}^{(GaN)}$, showing that this frequency can be much smaller than the resonance frequency of the particles.
In this case, the sideband of the higher-frequency particle resonates with the natural
mode of the lower-frequency particle, restoring spectral overlap that is otherwise
absent due to detuning. The combination of optimal phase ($\Delta\phi=\pi/2$) and
frequency matching ($\Omega= \Delta\omega_{\rm LO}$) thus maximizes
inelastic energy exchange between the modulated nanoparticles. On the contrary the inelastic channel is not sufficient to drive the transfer (Fig~\ref{Fig1:flux matching}).\\
\\
\textit{Directional Heat Flux.—} While elastic ($n=0$) energy exchange cancels at thermal equilibrium, the
inelastic channels ($n\neq 0$) still carry heat because the modulation shifts
photon frequencies to $\omega+n\Omega$, which probe different parts of the
thermal spectrum. As a result, the corresponding fluxes do not balance.  
Moreover, when the two particles have time-modulated polarizabilities with a
relative phase difference, the inelastic channels no longer compensate each
other, producing a net directional heat flux.

In a dipolar network at equilibrium, we can define the directional flux between two nodes $i$ and $j$ as
\begin{equation}
\mathcal{P}_{\rm dir} 
= \sum_{n=\pm 1} \int_0^\infty \frac{d\omega}{2\pi} 
\Big\langle P_{i\to j}^{(n)}(\omega) - P_{j\to i}^{(n)}(\omega) \Big\rangle .
\end{equation}
In the stationary reciprocal case, this contribution vanishes. On the other hand this is not the case anymore in a time-modulated system.
Using the inelastic transmission coefficient (\ref{Trans_inelastic}), the directional flux can be written as
\begin{equation}
\begin{split}
\mathcal{P}_{\rm dir} &= 
\sum_{n=\pm 1}\int_0^\infty \frac{d\omega}{2\pi}\,
2\hbar(\omega+n\Omega)\,n(\omega,T)\\
&\quad\times 
\frac{(\omega+n\Omega)^4}{c^4}\,
\mathrm{Tr}\!\Big[\mathbf{G}_{ij}(\omega+n\Omega)\,
\mathbf{G}_{ij}^\dagger(\omega+n\Omega)\Big] \\[2mm]
&\quad\times 
\Big(
\Im[\hat{\alpha}_{i,0}]\,\Im[\hat{\alpha}_{j,n}]
-\Im[\hat{\alpha}_{j,0}]\,\Im[\hat{\alpha}_{i,n}]
\Big),
\end{split}
\end{equation}
where $\hat{\alpha}_{k,0}$ and $\hat{\alpha}_{k,n}$ are the Floquet components of the
time-dependent polarizability $\alpha_k(t)$.
For a lossless harmonic modulation of polarizabilities, the
first-order sidebands contributions read
\begin{align}
\Im[\hat{\alpha}_{k,\pm1}]
= \pm \frac{\delta\alpha_k(\omega)}{2}\,\sin\phi_k,
\qquad k=i,j.
\end{align}
Introducing the complex modulation amplitude
\begin{equation}
\delta\hat{\alpha}_k(\omega)
=\frac{\delta\alpha_k(\omega)}{2}\,e^{i\phi_k},
\end{equation}
in the directional flux and keeping only terms
linear in the modulation amplitude yields
\begin{equation}
\begin{split}
\mathcal{P}_{\rm dir}
&\simeq 
2\int_0^\infty \frac{d\omega}{2\pi}\,
\hbar(\omega+\Omega)\,n(\omega,T) \\[1mm]
&\quad\times 
\Im[\delta\hat{\alpha}_i(\omega)]\;
\Im[\delta\hat{\alpha}_j(\omega+\Omega)] \\[1mm]
&\quad\times
\frac{(\omega+\Omega)^4}{c^4}
\,\mathrm{Tr}\!\Big[
\mathbf{G}_{ij}(\omega+\Omega)\,
\mathbf{G}_{ij}^\dagger(\omega+\Omega)
\Big]\,\sin(\Delta\phi),
\end{split}
\end{equation}
where $\Delta\phi = \phi_j - \phi_i$ is the phase difference between the two
modulations.
The simplification from Eq.~(12) to Eq.~(15) relies on a perturbative expansion in the modulation amplitude, where only terms linear in the modulation are retained. This approximation is valid in the weak-modulation regime $|\delta\alpha_i| \ll |\alpha_{i,0}|$, where higher-order contributions involving multiple modulation quanta scale with higher powers of $\delta\alpha_i$ and can therefore be neglected. In this regime, the dominant contribution to the directional flux arises from the interference between the static response and the first-order sideband fields, while higher-order interference processes become relevant only for stronger modulations beyond the perturbative regime.
The directional flux is therefore governed by phase-controlled interference between inelastic Floquet channels and vanishes for $\Delta\phi=0$, reaching its maximum for $\Delta\phi=\pi/2$.
\begin{figure}
	\centering
	\includegraphics[height=0.3\textwidth,angle=0]{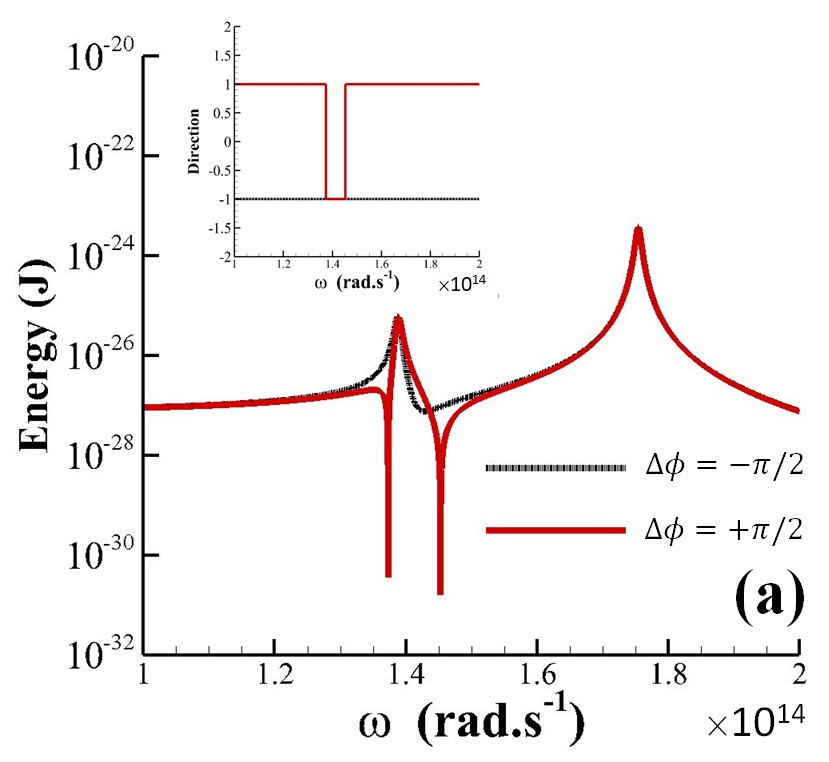}
           \includegraphics[height=0.3\textwidth,angle=0]{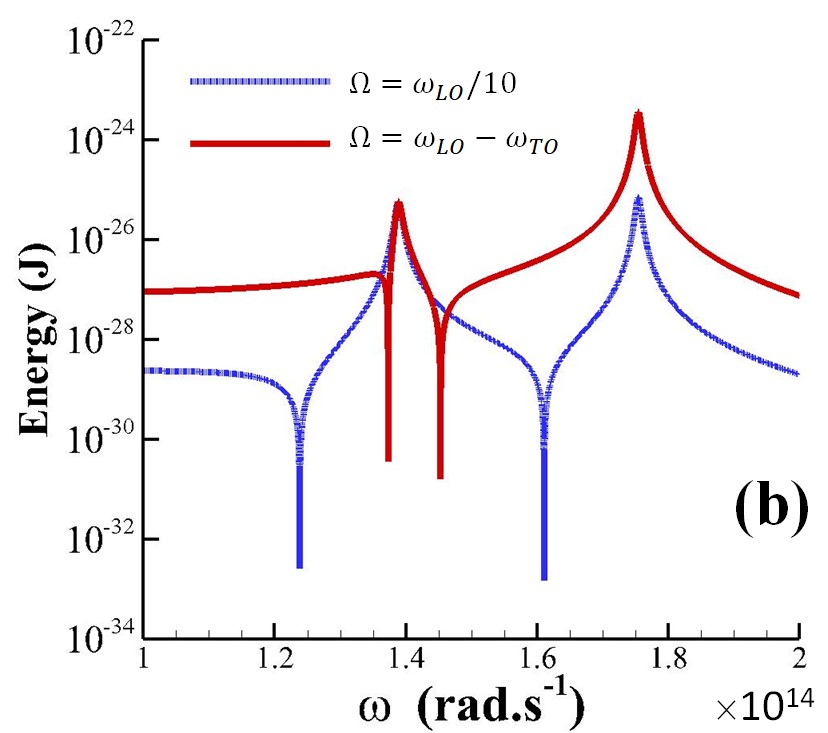}
	\caption{Directional energy exchange between two SiC nanoparticles at the same temperature $T=350\:K$ under a harmonic modulation (a) $\Omega=\omega_{\rm LO}^{(\mathrm{SiC})}-\omega_{\rm TO}^{(\mathrm{SiC})}$ of particles polarizability with a phase shift $\Delta\phi\equiv\phi_2-\phi_1=\pm \pi/2$. (b)  Harmonic modulation at two different frequencies with a phase shift $\Delta\phi= \pi/2$. Same geometrical and physical parameter as in Fig.1.Inset: direction of heat flux (equal to 1 (resp. -1), when the flux is in direction of particle 2 (resp. 1)).} \label{Fig3:Direction}
\end{figure} 
In Fig.~\ref{Fig3:Direction}, we plot the directional energy transfer between two SiC nanoparticles. The directional flux is maximized when the modulation frequency is comparable to the width  $\Delta\omega=\omega_{\rm LO}^{(\mathrm{SiC})}-\omega_{\rm TO}^{(\mathrm{SiC})}$ of the SiC Reststrahlen band, since in this regime one Floquet sideband overlaps a strong  phonon-polariton resonance while the opposite sideband remains off-resonant, resulting in maximal spectral asymmetry.
While the elastic contribution always flows from the hotter to the colder particle the inelastic flux can induce a transfer even at thermal equilibrium and the direction of flux (see inset of Fig~.\ref{Fig3:Direction}-a) is driven by the phase shift between the particles. 
\\
\\
\textit{Heat Flux Splitting.—} 
Consider a triplet of dipoles at positions $\mathbf{r}_i$, $\mathbf{r}_j$, and $\mathbf{r}_k$, with dipole~$i$ as the input and $j,k$ as outputs. Their induced dipoles satisfy
\begin{equation}
\mathbf{p}_i = \varepsilon_0 \sum_{j=1}^{3} [\mathbf{M}^{-1}]_{ij}\, \mathbf{E}_j^{\rm fl},
\end{equation}
where the component of the block matrix $\mathbf{M}$ are $\mathbf{M}_{ij} = \delta_{ij} \tilde{\alpha}_i \mathbf{I} + (1-\delta_{ij}) \tilde{\alpha}_i \, \mu_0 \omega^2 \mathbf{G}_{ij}$.
If only the output dipoles are harmonically modulated,
\begin{equation}
\alpha_l(t) = \alpha_{l,0} + \delta\alpha_l \cos(\Omega t + \phi_l), \quad l=j,k,
\end{equation}
first-order sidebands at $\omega \pm \Omega$ open inelastic channels. We emphasize that the phases $\phi_j$ and $\phi_k$ correspond to the phases of the external driving signals that modulate the optical properties of the nanoparticles. These phases are determined by the applied modulation fields rather than by the interparticle separation, allowing independent control even for nanoparticles separated by only a few hundred nanometers.
For real modulation amplitudes,
\begin{equation}
\Im[\hat{\alpha}_{l,\pm 1}] = \pm \frac{\delta\alpha_l}{2} \sin\phi_l, \qquad
\hat{\alpha}_{i,n}=0 \; (n\neq 0),
\end{equation}
so that the relative modulation phase $\Delta\phi = \phi_j - \phi_k$ controls the interference between the inelastic pathways.
We define the net power flux received by each output as
\begin{figure}
	\centering
	\includegraphics[height=0.3\textwidth,angle=0]{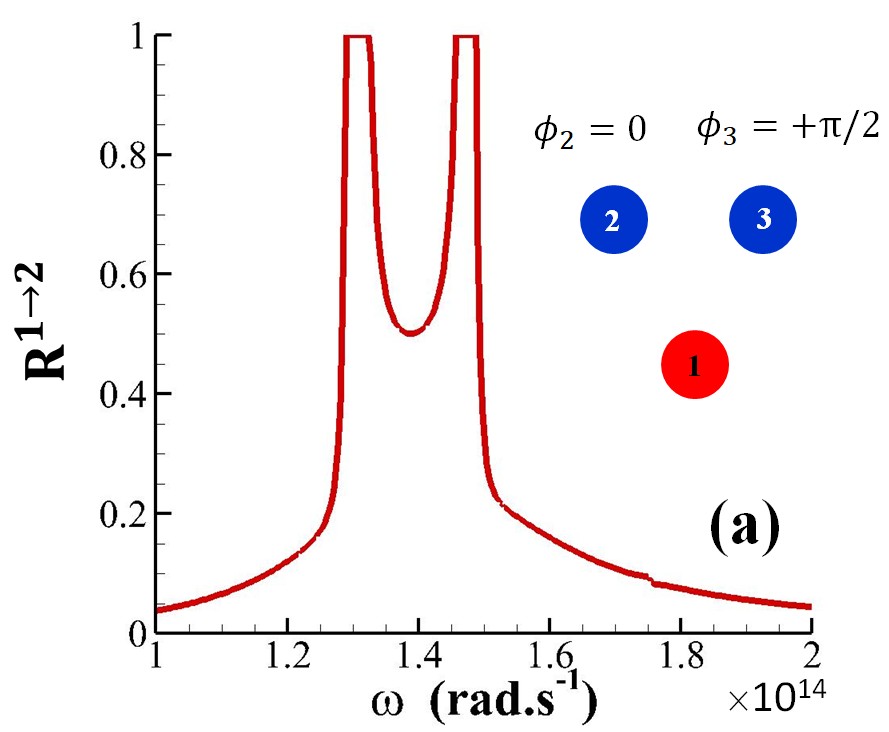}
           \includegraphics[height=0.3\textwidth,angle=0]{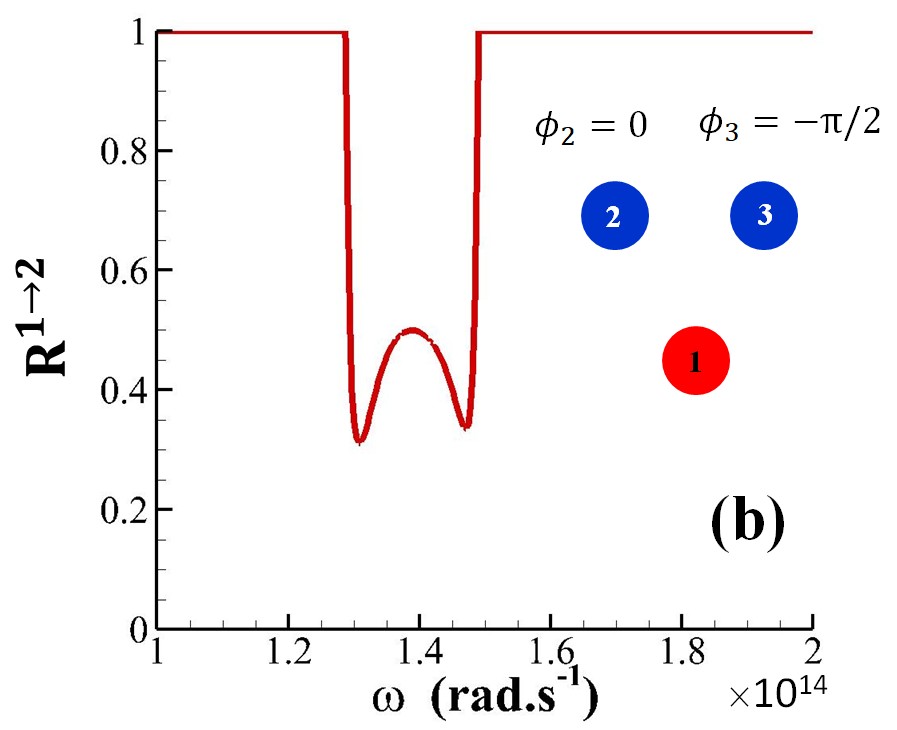}
	\caption{Phase-controlled splitting of radiative heat flux.
Splitting ratio $R^{1\to 2}$, defined from the net absorbed power, in a three-nanoparticle SiC--GaN--GaN network with $\delta\alpha=0.1$. The SiC nanoparticle (red) acts as a hot source at $T_1=400\,\mathrm{K}$, while the two GaN nanoparticles (blue) serve as colder outputs held at $T_2=T_3=300\,\mathrm{K}$. Panels (a) and (b) correspond to relative modulation phases $\Delta\phi=+\pi/2$ and $\Delta\phi=-\pi/2$, respectively. The GaN output nanoparticles are harmonically modulated at the frequency $\Omega=\omega_{\mathrm{LO}}^{(\mathrm{GaN})}-\omega_{\mathrm{TO}}^{(\mathrm{GaN})}$ with $\delta\alpha=0.1$.
}
 \label{Fig4:Splitting}
\end{figure} 
\begin{equation}
\mathcal{P}_{i \to j}(\omega) = \sum_{n=0,\pm1} (P_{i \to j}^{(n)}(\omega) - P_{j \to i}^{(n)}(\omega)),
\end{equation}
which accounts for actual measurable energy transfer, including both emission and backflow. Using this net flux, the flux splitting ratio is rigorously defined as
\begin{equation}
R^{i\to j}(\omega) = \frac{\mathcal{P}_{i \to j}(\omega)}{\mathcal{P}_{i \to j}(\omega) + \mathcal{P}_{i \to k}(\omega)}.
\end{equation}
By tuning $\Delta\phi$, the net flux can be redistributed between the two outputs while ensuring $0 \le R^{i\to j} \le 1$, thus avoiding unphysical values.
The definition generalizes straightforwardly to larger networks. For multiple input dipoles $\mathcal{I}$ and outputs $\mathcal{O}$, the modulation-controlled splitting from input $i\in\mathcal{I}$ to output $j\in\mathcal{O}$ becomes
\begin{equation}
R^{i\to j}(\omega) = \frac{\mathcal{P}_{i \to j}(\omega)}{\sum_{k\in \mathcal{O}} \mathcal{P}_{i \to k}(\omega)}.
\end{equation}
\begin{figure}
    \centering
    \includegraphics[height=0.3\textwidth,angle=0]{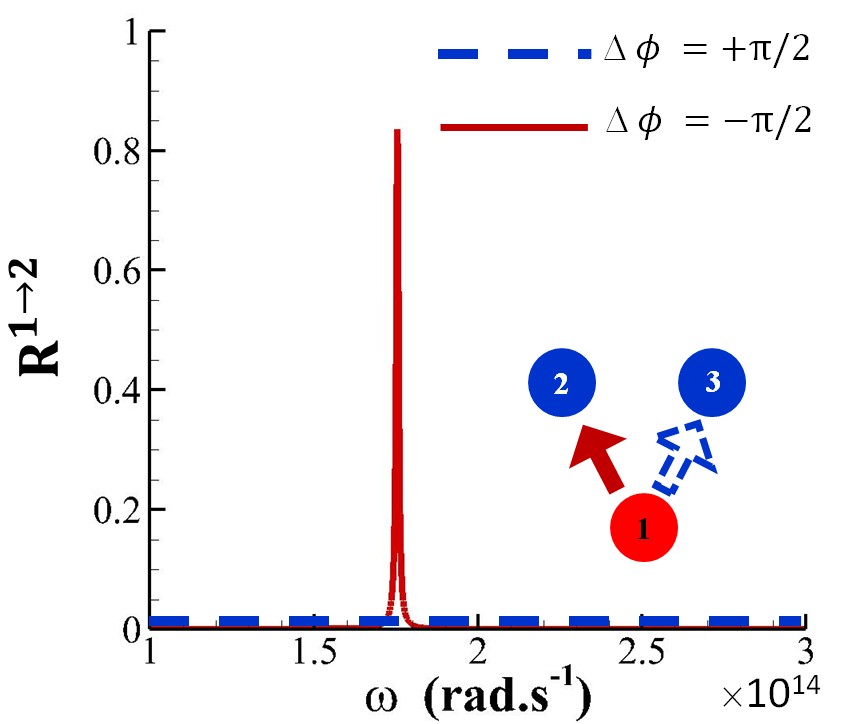}
   \caption{Phase-controlled splitting of radiative heat flux in a three-nanoparticle SiC--InSb--InSb network for a relative modulation phase $\Delta\phi=\pm\pi/2$. The SiC nanoparticle (red) acts as a hot source at $T_1=400\,\mathrm{K}$, while the two InSb nanoparticles (blue) serve as colder outputs at $T_2=T_3=300\,\mathrm{K}$. The InSb nanoparticles are harmonically modulated at a low frequency $\Omega=10^{10}\,\mathrm{rad\,s^{-1}}$ with a modulation amplitude $\delta\alpha=0.2$, which can be experimentally realized via piezoelectric actuation. The dielectric response of InSb is described by a Lorentz model~\cite{Palik} with parameters $\varepsilon_\infty=15.7$, $\omega_\mathrm{LO}=3.62\times10^{12}\,\mathrm{rad/s}$, $\omega_\mathrm{TO}=3.39\times10^{12}\,\mathrm{rad/s}$ and damping $\gamma=5.65\times10^{10}\,\mathrm{rad/s}$.The chosen modulation parameters correspond to conditions that can be experimentally realized via piezoelectric actuation.}
    \label{Fig4:Splitting_piezo}
\end{figure}
We demonstrate in Fig.~\ref{Fig4:Splitting} that a relative modulation phase between two output dipoles enables near-perfect splitting of radiative heat flux in a three-nanoparticle network composed of a hot SiC source coupled to two colder GaN nanoparticles. The particles form an equilateral triangular geometry in the near field, ensuring that heat transfer is governed by evanescent dipole–dipole interactions. The GaN outputs are harmonically modulated at a frequency 
$\Omega = \omega_{\mathrm{LO}}^{(\mathrm{GaN})} - \omega_{\mathrm{TO}}^{(\mathrm{GaN})}$, equal to the width of the GaN Reststrahlen band. This choice maximizes asymmetry between the Floquet sidebands, so that one sideband spectrally overlaps the GaN phonon–polariton resonance while the opposite sideband remains off-resonant. Consequently, the inelastic heat-transfer channels acquire strongly unbalanced amplitudes and efficiently interfere with the elastic pathway.  
The phase-dependent interference provides a direct, reversible control for directing energy flow. For a relative modulation phase $\Delta\phi=\pm\pi/2$, the inelastic contributions fully suppress the net heat current into one GaN nanoparticle while reinforcing it in the other, yielding complete thermal routing without modifying geometry or temperatures.  
This interference mechanism realizes a controllable thermal beam splitter at the level of net absorbed power. By associating the presence or absence of absorbed heat in a given output with logical states, the relative modulation phase acts as a control parameter enabling reconfigurable thermal logic and routing functionalities.
Notably, efficient control persists even when the modulation frequency is orders of magnitude below the resonance frequency, as the mechanism relies on phase-coherent interference rather than resonant excitation.
In Fig.~\ref{Fig4:Splitting_piezo} we present the heat-flux splitting at room temperature for a modulation frequency $\Omega = 10^{10}\,\mathrm{rad.s^{-1}}$. This modulation frequency  lies in a regime accessible via experimentally established techniques such as piezoelectric actuation~\cite{Hui,Chegel}, electro-optic modulation or coherent phononic excitation~\cite{Khalsa,Kusaba}. Patterned electrodes or localized strain fields allow independent tuning of the modulation phase for each nanoparticle, ensuring the experimental feasibility of the phase-controlled interference described here. We consider the same triangular geometry as before, with InSb nanoparticles replacing the GaN elements. The system exhibits phase-control of radiative heat transport: tuning the relative modulation phase $\Delta\phi$ modifies the interference between competing heat-transfer pathways.
A pronounced splitting, reaching nearly $85\%$, is observed for $\Delta\phi=-\pi/2$ near the surface phonon--polariton (SPhP) resonance of the SiC particle. In this configuration, interference between elastic and inelastic channels is constructive for the branch connecting particle~1 to particle~2, while being destructive for the branch toward particle~3, resulting in efficient routing of heat toward particle~2. 
In contrast, for $\Delta\phi=+\pi/2$, the interference pattern reverses. Near the SPhP resonance, destructive interference occurs simultaneously in both branches, quenching the net heat transfer rather than redirecting it and yielding an almost vanishing splitting ratio. Such a vanishing splitting ratio may arise either from symmetric suppression of radiative transport due to destructive interference or from negligible coupling far from resonance, and therefore does not necessarily imply redirection toward the opposite output. Slightly off resonance, where the elastic amplitude is reduced, interference can become relatively more favorable toward particle~3, allowing a fraction of the heat to be routed in this direction. However, far from resonance, both elastic and inelastic amplitudes vanish and the radiative flux toward both outputs disappears.
Microscopically, temporal modulation of the output particles at frequency $\Omega$ opens additional inelastic channels at $\omega\pm\Omega$, which coherently interfere with the elastic pathway. To leading order in the modulation amplitude, the scattering amplitudes from particle~1 to a modulated particle~$j$ read
\begin{equation}
\begin{split}
\mathcal{A}^{(0)}_{1j}(\omega)
&\propto
\alpha_1(\omega)\,\alpha_j^{(0)}(\omega)\,\mathbf{G}_{1j}(\omega),\\
\mathcal{A}^{(\pm1)}_{1j}(\omega)
&\propto
\alpha_1(\omega)\,\delta\alpha\,e^{\pm i\phi_j}\,
\mathbf{G}_{1j}(\omega\pm\Omega),
\end{split}
\end{equation}
where $\alpha_j^{(0)}$ is the static polarizability, $\delta\alpha$ the modulation amplitude, and $\phi_j$ the modulation phase. The scattered field is the coherent superposition of all pathways, yielding a spectral transmission probability
\begin{equation}
\mathcal{T}_{1j}(\omega)
\propto
\left|
\mathcal{A}^{(0)}_{1j}
+
\mathcal{A}^{(+1)}_{1j}
+
\mathcal{A}^{(-1)}_{1j}
\right|^{2}.
\end{equation}
The dominant interference term scales as
\begin{equation}
2\,\mathrm{Re}\!\left[
\mathcal{A}^{(0)}_{1j}
\mathcal{A}^{(\pm1)\,*}_{1j}
\right]
\propto
\cos\!\left(\phi_j+\theta_{1j}\right),
\end{equation}
where $\theta_{1j}$ is set by the complex polarizabilities and the Green tensor. These phase-dependent terms are strongly enhanced near the SiC resonance, where the elastic amplitude is maximal, and they determine whether heat is routed preferentially toward particle~2, toward particle~3, or is suppressed in both channels.

To conclude, we have shown that temporal modulation of material properties enables phase-controlled, radiative heat flux, achieving directional transport, reversible heat pumping, and tunable flux splitting—all without altering the structure. Coherent interference between elastic and inelastic channels provides a versatile mechanism for reconfigurable nanoscale thermal routing and logic operations.


\begin{thebibliography}{44}
\bibitem{Volokitin} A. I. Volokitin and B. N. J. Persson, Rev. Mod. Phys., \emph{Near-field radiative heat transfer and noncontact friction}, \textbf{79},1291 (2007).
\bibitem{Joulain} K. Joulain, J.-P. Mulet, F. Marquier, R. Carminati and J.-J. Greffet,  Surface Science Reports, \emph{Surface electromagnetic waves thermally excited: Radiative heat transfer, coherence properties and Casimir forces revisited in the near field}, 57, 59–112  (2005).
\bibitem{Cuevas} J. C. Cuevas and F. J. García-Vidal, ACS Photonics, \emph{Radiative Heat Transfer}, 5, 10, 3896–3915 (2018).
\bibitem{biehs1} S.-A. Biehs, R. Messina, P.~S. Venkataram, A.~W. Rodriguez, J.~C. Cuevas, and P. Ben-Abdallah, \emph{Near-field radiative heat transfer in many-body systems}, Rev. Mod. Phys. \textbf{93}, 025009 (2021).
\bibitem{Pendry} J. B. Pendry, J. Phys.: Condens. Matter, \emph{Radiative exchange of heat between nanostructures}, 11, 6621 (1999).
\bibitem{pba_prb_2010} P. Ben-Abdallah and K. Joulain, \emph{Fundamental limits for noncontact transfers between two bodies}, Phys. Rev. B \textbf{82}, 121419(R) (2010).
\bibitem{Miller_prl_2015} O. D. Miller, S. G. Johnson and A. W. Rodriguez, \emph{Shape-Independent Limits to Near-Field Radiative Heat Transfer}, Phys. Rev. Lett. \textbf{115}, 204302 (2015).
\bibitem{Messina_prb_2017} R. Messina, P. Ben-Abdallah, B. Guizal and M. Antezza, \emph{Graphene-based amplification and tuning of near-field radiative heat transfer between dissimilar polar materials}, Phys. Rev. B \textbf{96}, 045402 (2017).
\bibitem{Fan_prl_2018} H. Iizuka and S. Fan, \emph{Significant Enhancement of Near-Field Electromagnetic Heat Transfer in a Multilayer Structure through Multiple Surface-States Coupling}, Phys. Rev. Lett. \textbf{120}, 063901 (2018).
\bibitem{Rodriguez_prl2011} A. W. Rodriguez et al., \emph{Frequency-Selective Near-Field Radiative Heat Transfer between Photonic Crystal Slabs: A Computational Approach for Arbitrary Geometries and Materials}, Phys. Rev. Lett. \textbf{107}, 114302 (2011).
\bibitem{Cuevas_prl_2017} V.  Fernadez-Hurtado, F.J.  Garcia-Vidal, S. Fan and J. C. Cuevas, \emph{Enhancing Near-Field Radiative Heat Transfer with Si-based Metasurfaces}, Phys. Rev. Lett. \textbf{118}, 203901 (2017).
\bibitem{pba_prl_2014} P. Ben-Abdallah and S.-A. Biehs, Phys. Rev. Lett., \emph{Near-Field Thermal Transistor}, \textbf{112}, 044301 (2014).
\bibitem{Tang} G. Tang et al., \emph{Near-Field Energy Transfer between Graphene and Magneto-Optic Media}, Phys. Rev. Lett. \textbf{127}, 247401 (2021).
\bibitem{Song} J. Song, \emph{Magnetically Tunable Near-Field Radiative Heat Transfer in Hyperbolic Metamaterials}, Phys. Rev. Applied \textbf{13}, 024054 (2020).
\bibitem{Lingling} L. Fan et al., \emph{Nonreciprocal radiative heat transfer between two planar bodies}, Phys. Rev. B \textbf{101}, 085407 (2020).
\bibitem{Cuevas2} E. Moncada-Villa and J. C. Cuevas, \emph{Near-field radiative heat transfer between one-dimensional magnetophotonic crystals}, Phys. Rev. B 103, 075432 (2021).
\bibitem{pba2016} P. Ben-Abdallah, \emph{Photon thermal Hall effect}, Phys. Rev. Lett. \textbf{116}, 084301 (2016).
\bibitem{Liberal} J.~E. Vázquez-Lozano and I. Liberal, \emph{Incandescent temporal metamaterials}, Nat. Commun. \textbf{14}, 4606 (2023).
\bibitem{Fan1} Y. Renwen and S. Fan, \emph{Manipulating Coherence of Near-Field Thermal Radiation in Time-Modulated Systems}, Phys. Rev. Lett. \textbf{130}, 096902 (2023).
\bibitem{Wang1} H. Zhu et al., \emph{Enhancing far-field thermal radiation by Floquet engineering}, arXiv:2507.16688
\bibitem{Wang2} Y. Ren, H. Pan and J.-S. Wang, \emph{Clarification of Floquet--Enhanced Thermal Emission Through the Nonequilibrium Green's Function Formalism}, arXiv:2510.09300
\bibitem{Wang3} G. Tang and J.-S. Wang, \emph{Modulating near-field thermal transfer through temporal drivings: A quantum many-body theory}, Phys. Rev. B \textbf{109}, 085428 (2024).
\bibitem{Wang4} H. Pan, Y. Ren, G. Tang and J.-S. Wang, \emph{Asymmetry-induced radiative heat transfer in Floquet systems}, Phys. Rev. B \textbf{112}, L041401 (2025).
\bibitem{Hui} Y. Hui, J. Gomez-Diaz, Z. Qian et al., \emph{Plasmonic piezoelectric nanomechanical resonator for spectrally selective infrared sensing}, Nat Commun \textbf{7}, 11249 (2016).
\bibitem{Chegel} R. Chegel, \emph{Strain tuning of optical and thermoelectric properties of monolayer BAs}, Sci Rep \textbf{15}, 16227 (2025).
%
\bibitem{Khalsa} G. Khalsa, \emph{Ultrafast Control of Material Optical Properties via the Infrared Resonant Raman Effect}, Phys. Rev. X \textbf{11}, 021067 (2021).
\bibitem{Kusaba} S. Kusaba et al., \emph{Terahertz sum-frequency excitation of coherent optical phonons in the two-dimensional semiconductor} WSe$_2$, Appl. Phys. Lett. \textbf{124}, 122204 (2024).

\bibitem{Alcazar} L.~J. Fernandez-Alcázar, H. Li, M. Nafari, and T. Kottos, \emph{Implementation of Optimal Thermal Radiation Pumps Using Adiabatically Modulated Photonic Cavities}, ACS Photonics \textbf{8}, 2973 (2021).
\bibitem{Fan2} R. Yu and S. Fan, \emph{Time-modulated near-field radiative heat transfer}, PNAS \textbf{121}, e2401514121 (2024).
\bibitem{Fan3} S. Buddhiraju, W. Li, and S. Fan, \emph{Photonic Refrigeration from Time-Modulated Thermal Emission}, Phys. Rev. Lett. \textbf{124}, 077402 (2020).
\bibitem{arXiv:2510.19378}, R. Messina and P. Ben-Abdallah, \emph{Many-Body Floquet Theory for Radiative Heat Transfer in Time-Modulated Systems}, Phys. Rev. B \textbf{113}, 035404 (2026). 
\bibitem{Zhao}  C.-L. Zhao, Z.Torbatian, S.-H. Yang, Y. Zhang, H.-L. Yi, M. Antezza, D. Novko and C.-W. Qiu, \emph{Unconventional Thermophotonic Charge Density Wave}, Phys. Rev. Lett. \textbf{133}, 066902 (2024).
\bibitem{pba2011}P. Ben-Abdallah, S.-A. Biehs, and K. Joulain, \emph{Many-Body Radiative Heat Transfer Theory}, Phys. Rev. Lett. \textbf{107}, 114301 (2011). 
\bibitem{messina2013}R. Messina, M. Tschikin, S.-A. Biehs, and P. Ben-Abdallah, \emph{Fluctuational-electrodynamic theory and dynamics of heat transfer in multiple dipolar systems}, Phys. Rev. B \textbf{88}, 104307 (2013).


\bibitem{Palik}\emph{Handbook of Optical Constants of Solids}, edited by E. Palik (Academic Press, New York, 1998).


\end{thebibliography}
\end{document}